# Functional analysis of spontaneous cell movement under different physiological conditions

Hiroaki Takagi[1,2,3,*], Masayuki J. Sato[1,3], Toshio Yanagida[1], Masahiro Ueda[1,3]

1. *Laboratories for Nanobiology, Graduate School of Frontier Biosciences, Osaka University, 1-3 Yamadaoka, Suita, Osaka, 565-0871, Japan.*
2. *Department of Physics, School of Medicine, Nara Medical University, Nara, 634-8521, Japan.*
3. *JST, CREST, 1-3 Yamadaoka, Suita, Osaka, 565-0871, Japan.*

[*] To whom correspondence should be addressed.
E-mail: takagi@naramed-u.ac.jp

**Abstract**

Cells can show not only spontaneous movement but also tactic responses to environmental signals. Since the former can be regarded as the basis to realize the latter, playing essential roles in various cellular functions, it is important to investigate spontaneous movement quantitatively at different physiological conditions in relation to a cell's physiological functions. For that purpose, we observed a series of spontaneous movements by *Dictyostelium* cells at different developmental periods by using a single cell tracking system. Using statistical analysis of these traced data, we found that cells showed complex dynamics with anomalous diffusion and that their velocity distribution had power-law tails in all conditions. Furthermore, as development proceeded, average velocity and persistency of the movement increased and as too did the exponential behavior in the velocity distribution. Based on these results, we succeeded in applying a generalized Langevin model to the experimental data. With this model, we discuss the relation of spontaneous cell movement to cellular physiological function and its relevance to behavioral strategies for cell survival.

**Introduction**

Cell movement is composed of multiple dynamical processes such as surface attachment and detachment cycles, development and collapse of filopodia, movement of the cell body center, and maintenance of cell morphology. In these processes, the motile apparatus such as the cytoskeleton or a motor molecule and the many related signaling molecules systematically coordinate to achieve proper function [1]. Cell movement can be distinguished between spontaneous movements and tactic responses to environmental signals. Spontaneous cell movement is a random motion under no external guiding cues, which accompanies large fluctuations in the dynamical localizations of corresponding molecular components in order to coordinate function. Tactic behaviors are achieved by biasing the cell movement in a sensitive and stable manner in response to environmental signals [2,3], thus playing an essential role in various cellular functions. Consequently, it is important to quantify the existing fluctuations in cell motion dynamics and identify its control mechanism at different physiological conditions in order to clarify the physiological meaning of spontaneous cell movement. For this purpose, we adopted *Dictyostelium discoideum* (Fig. 1) as a model to quantify spontaneous movement. *Dictyostelium* cells are a well established model for the study of amoeboid movement and tactic responses, and also for development [3,4,5]. These cells have a simple life cycle: they grow as separate, independent cells while ingesting bacteria but interact to form multicellular structures (slugs) when challenged by adverse conditions such as starvation. Since the vegetative and developmental stages are completely independent and during starvation the cell drastically changes its physiological state, we measured a series of spontaneous cell movements by using a single cell tracking system during transient developmental processes. Through statistical analysis of these data, we present the characteristics of the dynamics and the way of their developmental change. We then investigated the applicability of the generalized Langevin model to the experimental data. Finally, we discuss the relevance of spontaneous cell movement to cellular physiological functions with regards to behavioral strategies for cell survival.

**Materials and Methods**

During our experiments, we made the cell density low to remove any explicit cell-cell interaction effects and thus established a uniform environment (typical experimental conditions for vegetative and 5.5 hours starved cells are shown in Video S1 and S2, respectively). Experimental procedures are as follows [3]:

(1) Cell preparation. *Dictyostelium discoideum*, Ax2 cells (wild type) were grown at 21°C in HL5 medium supplemented with $5\,ng\cdot ml^{-1}$ vitamin B12 and $100\,ng\cdot ml^{-1}$ folic acid. After washing and resuspending in development buffer (DB; 10 mM Na/K PO$_4$, 2 mM MgSO$_4$, 0.2 mM CaCl$_2$, pH 6.5) at a density of $5.0\times10^6\,cells\cdot ml^{-1}$, cells were placed in a plastic culture dish (35 mm culture dish, Iwaki) and maintained at 21°C. After 1 hour, pulse stimulation with 100 nM cAMP commenced at 6 min intervals for up to 2.5, 4, or 5.5 hours on a rotating shaker (SL3D, SeouLin Bioscience). After washing, the cell density was adjusted appropriately to reduce cell-cell interaction (about 80 $cells \cdot mm^{-2}$). The cell suspension was then injected into an observation chamber. After settling for 10 min, cell behavior was observed. All experiments were performed at 21$^\circ$C room temperature.

(2) Chamber. The chamber consisted of two metal chambers (Attofluor Cell Chamber, A-7816, Molecular Probes) and a coverslip (circular type, diameter is 25 mm, MATSUNAMI). Used coverslips were first sonicated in 70 % EtOH for 15 min and dried. Coverslips were sandwiched between two metal chambers. Before injecting the cell suspension, coverslips were washed with DB.

(3) Microscopy and cell motility analysis. The cells in the chamber were observed with an Olympus IX-71 inverted microscope capable of producing phase contrast optics. The behavior of the cells was recorded with a cooled CCD camera (MicroMax, Princeton Instruments inc.) and MetaMorph software (Molecular Devices). Cell images were acquired at 1 sec intervals for 40 min. To analyze the motile activities, individual cells were followed automatically by using lab developed software. Their geometrical center positions were determined in x,y-coordinates (Fig. 1(c),(d)). From the positional changes, the motile properties were analyzed. Experiments were done at the vegetative state (denoted as 0hr) and at several transient cell developmental periods (denoted as 1hr, 2.5hr, 4hr and 5.5hr of the starved period, respectively).

(4) Statistical analysis. To analyze the motile activities of *Dictyostelium* cells in all conditions, we calculated a series of statistical quantities as follows: mean square displacement (*MSD*) of cell trajectories and its logarithmic derivative (*dln(MSD)/dln(t)*) (Fig. 2(a),(b)), velocity distribution (x-component, Fig. 2(c)), dependence of the kurtosis of the displacement distribution on time (x-component, Fig. 2(d)), velocity autocorrelation (Fig. 2(e)), dependence of the average turn angle on time (Fig. 2(f)), and dependence of the average turn rate on velocity (Fig. 2(g)). By setting a positional vector on the *i*-th cellular trajectory at time *t* as $\mathbf{p_i}(t) = (x_i(t), y_i(t))$, the *MSD* is defined as $MSD(t) = <|\mathbf{p_i}(t'+t) - \mathbf{p_i}(t')|^2>_{t',N}$, where $<>_{t',N}$ is the temporal and ensemble average of all the trajectories. Velocity is defined as $\mathbf{v_i}(t) = (\mathbf{p_i}(t + \Delta t) - \mathbf{p_i}(t))/\Delta t$ with $\Delta t = 1 \sec$. The kurtosis of the displacement distribution (x-component) is defined as

$$K(t) = \frac{<d_{ix}(t',t) - \mu(t)>^4_{t',N}}{\sigma(t)^4} - 3,$$

where $d_{ix}(t',t)$ is the x-component of $\mathbf{d_i}(t',t) = \mathbf{p_i}(t'+t) - \mathbf{p_i}(t')$, $\mu(t) = <d_{ix}(t',t)>_{t',N}$, and $\sigma(t) = <(d_{ix}(t',t) - \mu(t))^2>_{t',N}$. Kurtosis is a measure of whether the target distribution is peaked ($K(t) > 0$) or flat ($K(t) < 0$) relative to the Gaussian distribution. Velocity autocorrelation is defined as $C(t) = <\mathbf{v_i}(t') \cdot \mathbf{v_i}(t'+t)>_{t',N}$. The average turn angle is defined as $\theta(t) = <|\cos^{-1}(\frac{\mathbf{d_i}(t',t) \cdot \mathbf{d_i}(t'+t,t)}{|\mathbf{d_i}(t',t)||\mathbf{d_i}(t'+t,t)|})|>_{t',N}$. The average turn rate is defined as $\Delta\theta(v) = <|\cos^{-1}(\frac{\mathbf{v_i}(t) \cdot \mathbf{v_i}(t+\Delta t)}{|\mathbf{v_i}(t)||\mathbf{v_i}(t+\Delta t)|})|>_{t,N}$ with $\Delta t = 1 \sec$. Note that we confirmed that cell movement was not biased toward a specific direction in all

conditions studied (see Figure S1).

**Results**

By statistical analysis, we identified the characteristics of spontaneous cell movement. From the *MSD* of the cell trajectories and its logarithmic derivative (*dln(MSD)/dln(t)*) shown in Fig. 2(a,b), we found that *dln(MSD)/dln(t)* did not converge to unity, and in all conditions except for vegetative cells, there were three timescales in the cell movement: several seconds, a couple of minutes, and around ten minutes. As development proceeded, the motion became proportional to $t^{2H}$ (*H* is Hurst exponent, here $H = 0.62 \sim 0.83$) in the *2*nd timescale motion. These properties were confirmed by examining the dependence of the average turn angle on time as shown in Fig. 2(f). In addition, the velocity distribution was non-Gaussian for all conditions (Fig. 2(c)) and the kurtosis of the displacement distribution took positive values at least during the *1*st and *2*nd timescale motion indicating a peaked distribution relative to the Gaussian (Fig. 2(d)). Also the temporal correlation of the cellular velocity was not a single exponential function. In fact, as the development proceeded, the correlation became stronger so that its functional form was fitted to a two exponential function (Fig. 2(e)). In contrast, when applying the Ornstein-Uhlenbeck process, which is the simplest description for persistent random motion [6], *dln(MSD)/dln(t)* converged to unity, the velocity distribution was Gaussian, and the velocity autocorrelation was single exponential. Putting these results together, we confirmed that the cell movement is not simply described by a persistent random walk but contains complex dynamics with anomalous super-diffusion processes [7] and its persistency and velocity increase with development (see also Figure S1).

Here we focused on the cell movement with the *1*st and *2*nd timescale dynamics and investigated the mechanism used to produce the presented characteristics. From the dependence of the average turn rate on velocity shown in Fig. 2(g), anti-correlative behaviors were confirmed in all conditions studied [4,8]. This characteristic can be described by the Langevin equation with an angular coordinate for a two dimensional Ornstein-Uhlenbeck process, $|d\theta/dt|=|f_\theta|/mv$ ( $f_\theta$ is zero-mean Gaussian white noise, $m$, $\theta$ and $v$ are mass, turn angle and velocity of the cell, respectively), which supports the applicability of a Langevin-type phenomenological model to the cell movement [8]. However, the model is not a simple Ornstein-Uhlenbeck process as mentioned above, so we next estimated the form of the distribution and autocorrelation of velocity in detail to modify the model. For this reason, we show the results of the vegetative and 5.5hr starved cells in comparison. The velocity distributions (x-component) and fitting curves are shown in Fig. 3(a). Both groups of data had power-law tails and could not be fitted to either single Gaussian or single exponential functions, although the exponential tendency was higher in 5.5hr starved cells (we confirmed the tendency in all conditions studied). We also show velocity autocorrelations and fitting curves in Fig. 3(b). A two exponential function better fit to both data than a power-law function (vegetative: $\chi^2$ =34.6 (two exponential), 475.9 (power-law); 5.5hr starved: $\chi^2$ =7140 (two exponential), 53869 (power-law)). Their characteristic timescales were (11 sec, 1.8min) for vegetative and (5.2 sec, 3.8 min) for 5.5hr starved cells, respectively. Furthermore, we calculated the dependence of acceleration and fluctuation on velocity in cells by defining parallel and perpendicular components to the moving direction (Fig. 3(c),(d)). Note that in the case of the Ornstein-Uhlenbeck process, acceleration from the parallel component had a linear dependency on the velocity and fluctuation from either component was independent of the velocity, which satisfied fluctuation-dissipation relation. In the cell movement, however, acceleration from the parallel component and fluctuation from either component have a nonlinear dependency on the velocity in all conditions studied.

From these results, we applied the generalized Langevin model to the cell movement. Observed characteristics of velocity distribution and autocorrelation were reproducible by a state-dependent, additive noise and a memory term in the velocity function. With nonlinearities in acceleration and fluctuation, the model is described as [9]

$$\frac{d\mathbf{v}(t)}{dt} = -\beta(v(t))\mathbf{v}(t) + \alpha\mathbf{V}(t) + \sigma(v(t))\mathbf{h}(t),$$

$$\frac{d\mathbf{V}(t)}{dt} = \alpha\mathbf{v}(t) - \gamma\mathbf{V}(t),$$

where $\mathbf{v}(t)$ is the velocity of a cell, $\mathbf{V}(t)$ is the memory of the velocity defined as $\mathbf{V}(t) = \alpha \int_{-\infty}^{t} e^{-\gamma(t-t')} \mathbf{v}(t')dt'$ ($\alpha$ is the memory rate, $\gamma$ is the memory decay rate), $\beta(v(t))$ is the velocity decay rate with velocity dependency, and $\mathbf{h}(t)$ and $\sigma(v(t))$ is the noise of a cell and its strength, respectively. We note that $\sigma(v(t))$ has a parallel ($\sigma_{\parallel}(v(t))$) and a perpendicular component ($\sigma_{\perp}(v(t))$) to $\mathbf{v}(t)$. Here we assume that the source noise is zero-mean Gaussian white noise. To produce the presented nonlinear dependencies and power-law tails in the velocity distribution, we employ a cubic polynomial both for $\beta(v(t))$ and $\sigma(v(t))$, so that the model contains additive and multiplicative noises. Because the model is highly nonlinear, we can only approach the behaviors of the model numerically. Numerical integration of this stochastic differential equation was done by using a second order Runge-Kutta method, Heun scheme to realize Stratonovich-type integrals. We obtained theoretical trajectories of cellular motions from numerical simulations and calculated the corresponding statistical quantities. By searching the wide range of parameters and fitting the simulation results to the experimental ones, we can estimate corresponding values of parameters and functional forms. For comparison, we show the corresponding results of vegetative (0hr) and fully starved cells (5.5hr) (Fig. 4). The values of the parameters obtained were

$\alpha$ =0.34 ($s^{-1}$), $\gamma$ =0.095 ($s^{-1}$) at the vegetative state, and $\alpha$ =0.098 ($s^{-1}$), $\gamma$ =0.029 ($s^{-1}$) at the starved state. The functional form of $\beta(v)$, $\sigma_{\parallel}(v)$ and $\sigma_{\perp}(v)$ are shown in Table 1. By using this model, although there are some discrepancies between the experiment and simulation results in Fig. 4, almost all the characteristics of the cell movement presented above can be reproduced. We found that with development, the characteristic time of the memory became longer while the rate constant of the memory and the strength of the multiplicative noise became smaller.

**Discussion**

Here we investigated the dynamics of spontaneous cell movement at different developmental periods of *Dictyostelium* cells. The results can be summarized in three main points. First, the cell movement showed complex dynamics with anomalous diffusion where several characteristic timescales existed. Second, the velocity distribution had power-law tails in all conditions studied, and as development proceeded, highly correlated movement appeared to be dominant. According to this correlation, the velocity and persistency of the cell movement increased as too did the exponential behavior of the velocity distribution. Third, by applying the Langevin model to the cell motion dynamics, we could quantify several physical terms including noise, memory and decay rate of the cellular velocity. We determined the corresponding model parameters through numerical calculations and fitting the experimental data, finding that the functional forms of the decay rate and noise of the cellular velocity are nonlinear and the existence of additive and multiplicative noises and memory in the velocity dynamics. As for the physiological origins of the two time scales in the temporal correlation analysis, a possible candidate for the shorter period (5.2 sec) is filopodia formation dynamics while the longer period (3.8 min) may be due to the persistence time of the directional movement of the cell. Temporal analysis of the angular dynamics supports this view. As for the origins of the memory term and functional differences in Table 1, they closely relate to the formation of the cell's polarity, which enable the cells to realize stable movement and tactic behaviors. Regarding the origins of the noise term, the molecular system producing the additive and multiplicative noises remains for future studies.

These dynamic characteristics of spontaneous cell motion have possible physiological meanings in view of the behavioral strategies for cell survival. In vegetative cells, Levy-type self-avoiding walk is possibly the optimal strategy for foraging [10] while as development proceeds, starved cells assemble to make multicellular organism slugs such that correlated walking (and possibly the *3*rd timescale motion around ten minutes) is advantageous in order to make cells adjoin. The ability to produce these different types of movements and the interchanging between them can contribute to evolutionary advantages for the cells. As for the exponential property of the velocity distribution in starved cells, it can be consistent with the energy balance and partitioning necessary for the cell survival in a no energy resource environment [11]. Thus, spontaneous cell movement is not only necessary for rich dynamics in itself but is also relevant to the behavioral strategies for cell survival. These studies should lead to a deeper understanding of the underlying mechanism and also the physiological meaning relating to cellular spontaneous fluctuations.

**Acknowledgments**

We thank Tatsuo Shibata for valuable discussions and Peter Karagiannis for revising the manuscript.

**Figure legends & Table**

**Figure 1**

*Dictyostelium* cell in the vegetative and starved state, along with the corresponding cellular trajectories. (a) and (c) are vegetative, (b) and (d) are starved (for 5.5hr). Scale bar is $10\,\mu m$. In each case, we plot the trajectories of 20 cells for 40 minutes. All original cell positions are set to the axis origin. As development proceeds, cell polarity also develops.

**Figure 2**

Analyzed motile properties of *Dictyostelium* cells for all the experimental conditions. (a),(b): Mean square displacement (*MSD*) and its logarithmic derivatives are shown in a log-log plot and semi-log plot, respectively. (c):

Velocity distribution for the x axis is shown in a semi-log plot. Velocity per second was calculated and shown in minutes. (d): Dependence of the kurtosis of the x-axis component displacement distribution on time. (e): Autocorrelation function of velocity. (f): Dependence of the average turn angle on time. (g): Dependence of the average turn rate on velocity. In each developmental period, cellular tracks for 40 minutes were sampled over 100 cells to calculate all the above quantities. Standard errors are plotted together in (a),(e),(f),(g) (most error bars are smaller than the symbols).

**Figure 3**

Velocity distribution, autocorrelation and dependence of mean and standard deviation (*SD*) of acceleration on velocity. (a),(b): (a) Velocity distribution for the x-axis and (b) velocity autocorrelation of vegetative and starved cells (5.5hr) are shown in a log-log plot. Exponential (two exponential in (b)) and Gaussian fitting curves are plotted together. Power-law lines are guides for the reader. (c),(d): Dependence of mean and standard deviation (*SD*) of acceleration on velocity. Parallel and perpendicular components to the moving direction are calculated. (c) vegetative cells, (d) starved cells (5.5hr).

**Figure 4**

Simulation results corresponding to vegetative and starved cells. Simulated velocity distribution ((a),(b)), velocity autocorrelation ((c),(d)) and dependence of mean and standard deviation of acceleration on velocity in parallel and perpendicular components to the moving direction ((e),(f)) in the case of vegetative and starved cells (5.5hr). Corresponding experimental results are plotted together.

**Table1** : The functional form of $\beta(v)$, $\sigma_{\parallel}(v)$ and $\sigma_{\perp}(v)$.

|  | $\beta(v)(s^{-1})$ | $\sigma_{\parallel}(v)(\mu m/s^{2/3})$ | $\sigma_{\perp}(v)(\mu m/s^{2/3})$ |
|---|---|---|---|
| 0hr | $2.9 + 2.7v^2$ | $0.040 + 1.33v + 1.58v^2$ | $0.040 + 0.35v + 0.75v^2$ |
| 5.5hr | $0.50 + 1.7v^2$ | $0.050 + 0.80v + 0.60v^2$ | $0.050 + 0.40v$ |

* These are results from the numerical simulation fitted to the experimental data.

**Figure S1**

Dependence of average cellular velocity and directness on developmental period. The definition of cellular velocity is described in the main text, and the temporal and ensemble average of the velocity in each developmental period is plotted. Directness of *i-th* cell is defined as $\cos\theta_i$, where $\theta_i$ is the angle between the direction of displacement of *i-th* cell and the direction of horizontal axis. Average directness of the cell population was calculated as $<\cos\theta_i>_N$, where $<>_N$ is the ensemble average. Thus, in a randomly moving population of cells the value takes zero, while in a population of cells moving to a specific direction (horizontal axis) the value takes unity. Standard errors for average velocity and standard deviations for average directness are plotted together.

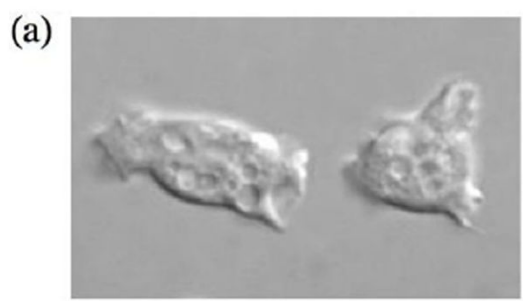 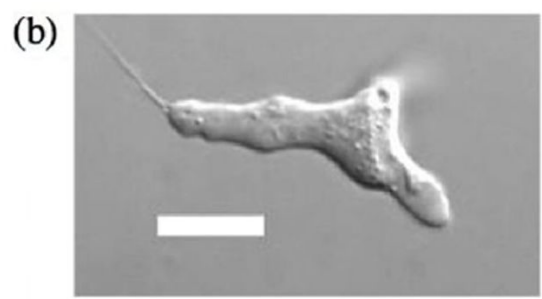
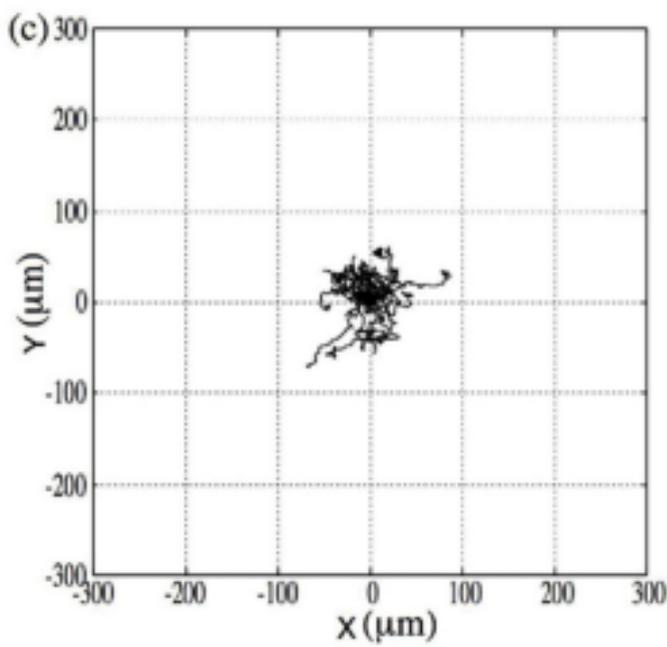 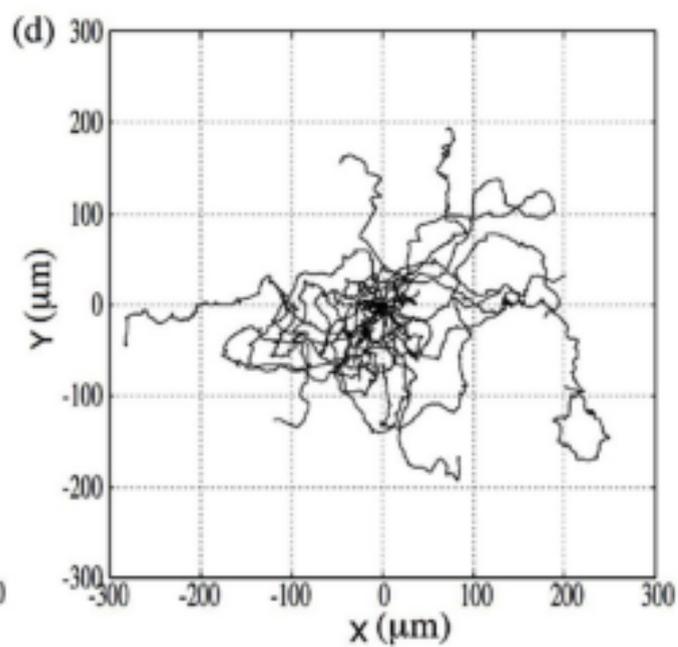

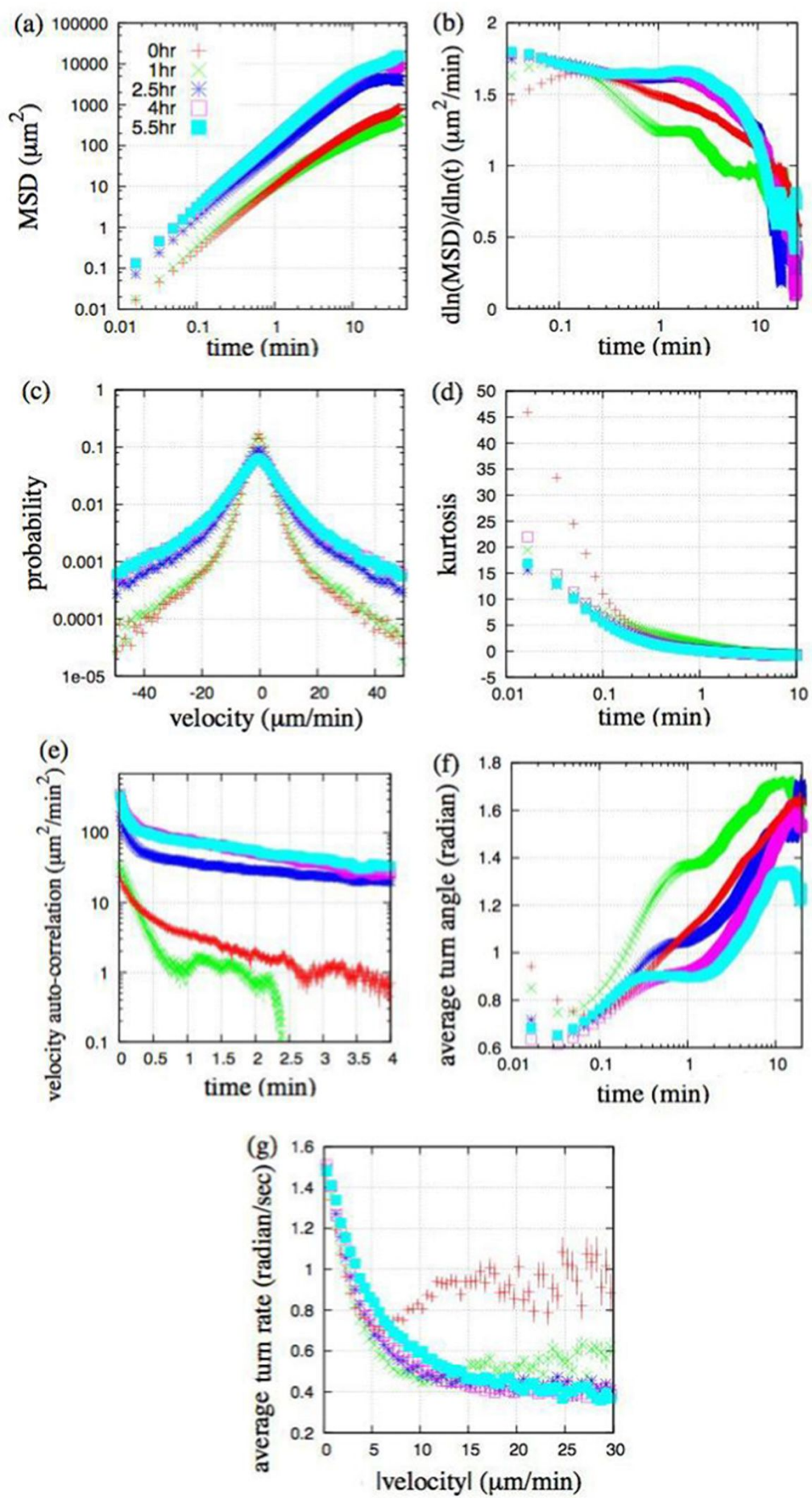

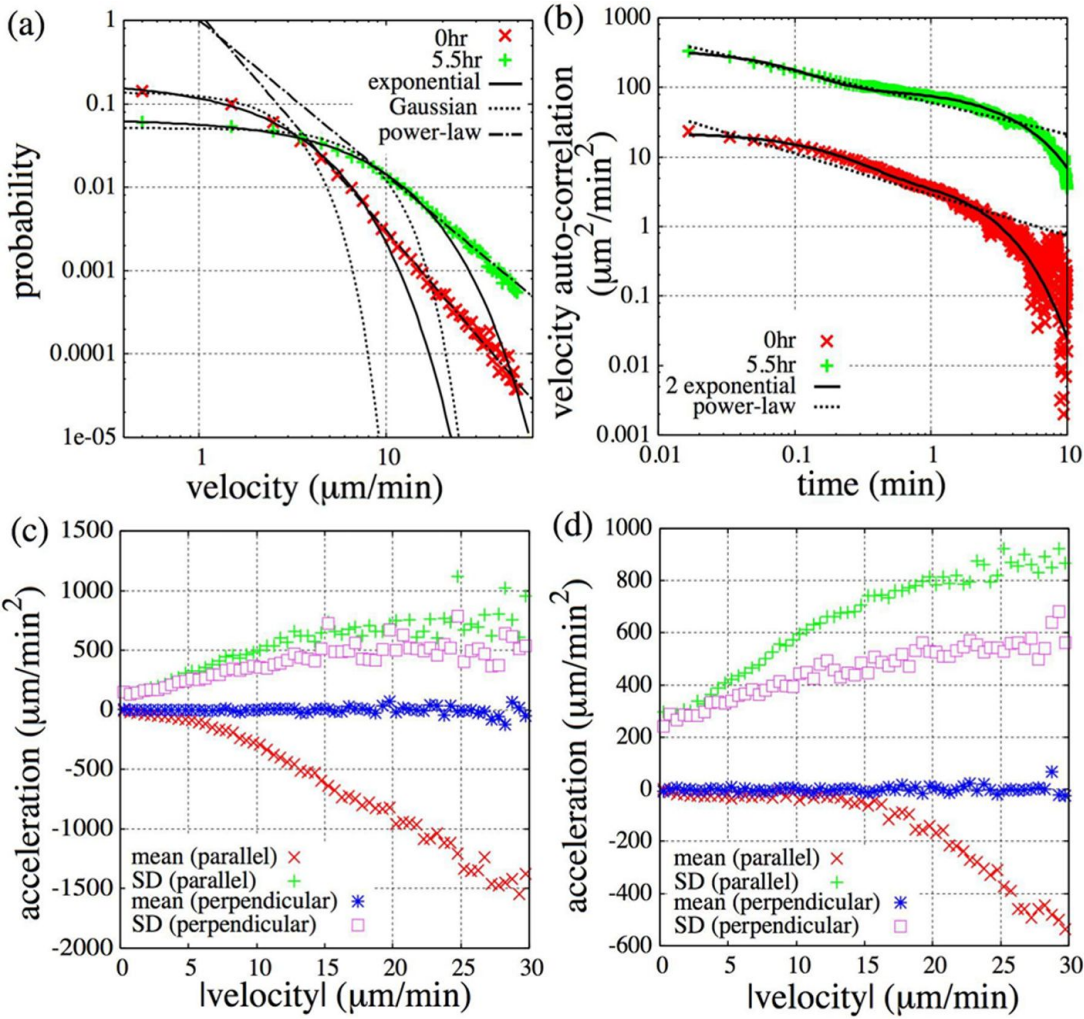

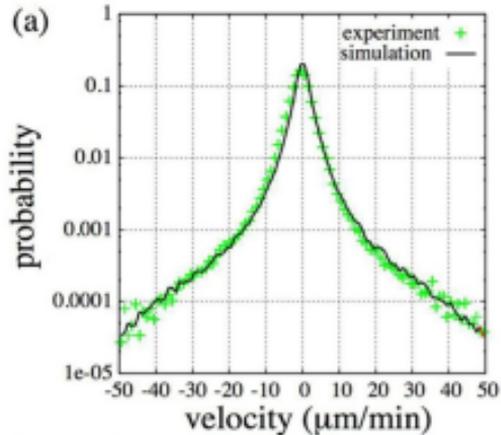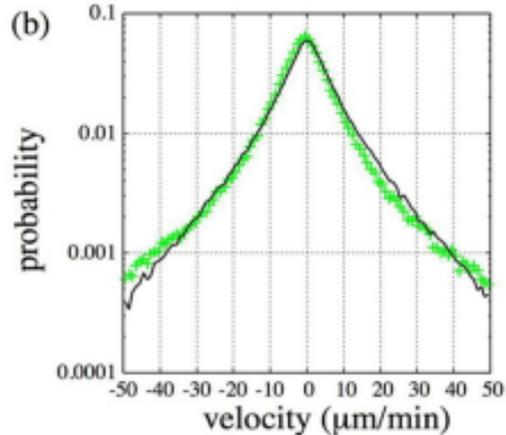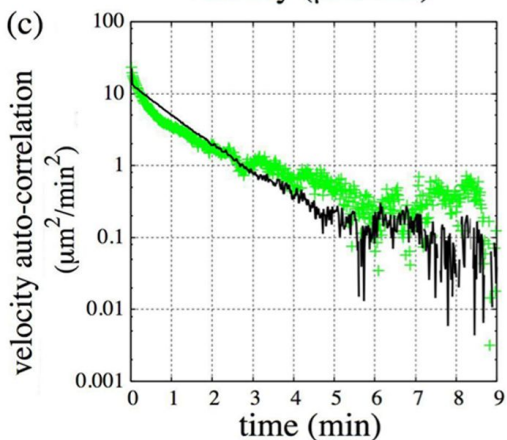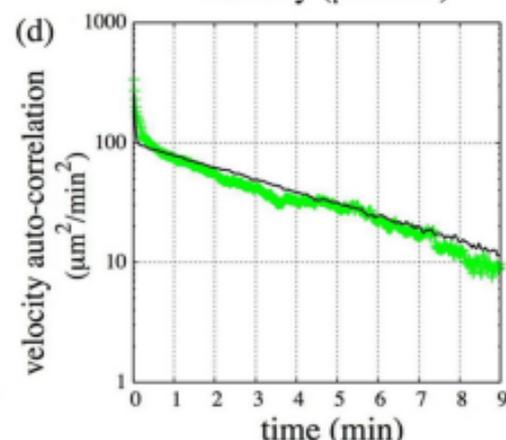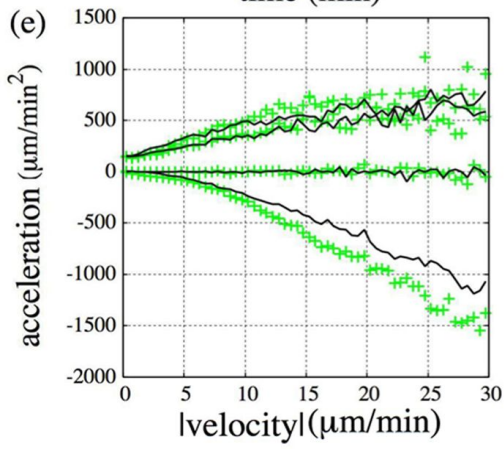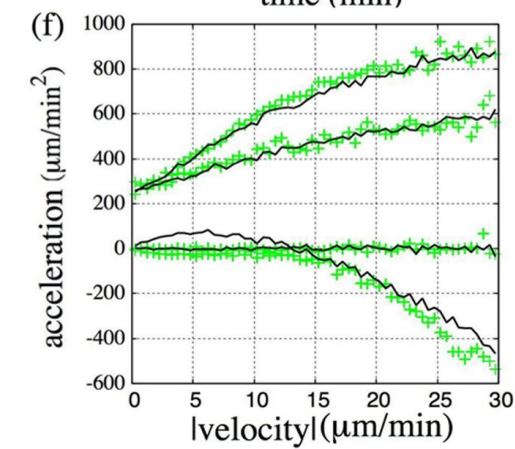